\RequirePackage{fix-cm}
\documentclass[smallextended]{svjour3}       
\smartqed  
\usepackage{graphicx}
\usepackage{multirow}
\usepackage{amssymb}
\usepackage{color}
\usepackage{makecell}
\begin{document}

\title{F3RNet: Full-Resolution Residual Registration Network for Deformable Image Registration
}

\titlerunning{Full-Resolution Residual Registration Network}        

\author{Zhe Xu         \and
        Jie Luo \and Jiangpeng Yan \and Xiu Li \and Jagadeesan Jayender 
}


\institute{Zhe Xu \at
              Shenzhen International Graduate School, Tsinghua University, Shenzhen 518055, China \\
              Brigham and Women's Hospital, Harvard Medical School, Boston 02115, USA          
           \and
           Jie Luo \at
              Brigham and Women's Hospital, Harvard Medical School, Boston 02115, USA \\
         \and
           Jiangpeng Yan, Xiu Li \at
            Shenzhen International Graduate School, Tsinghua University, Shenzhen 518055, China \\
            \email{li.xiu@sz.tsinghua.edu.cn} 
                 \and
           Jagadeesan Jayender \at
            Brigham and Women's Hospital, Harvard Medical School, Boston 02115, USA \\
            \email{jayender@bwh.harvard.edu} 
}
\date{Received: date / Accepted: date}

\maketitle

\begin{abstract}
Purpose: Deformable image registration (DIR) is essential for many image-guided therapies. Recently, deep learning approaches have gained substantial popularity and success in DIR. Most deep learning approaches use the so-called mono-stream ``high-to-low, low-to-high” network structure, and can achieve satisfactory overall registration results. However, accurate alignments for some severely deformed local regions, which are crucial for pinpointing surgical targets, are often overlooked. Consequently, these approaches are not sensitive to some hard-to-align regions, e.g., intra-patient registration of deformed liver lobes.

Methods: We propose a novel unsupervised registration network, namely Full-Resolution Residual Registration Network (F3RNet), for deformable registration of severely deformed organs. The proposed method combines two parallel processing streams in a residual learning fashion. One stream takes advantage of the full-resolution information that facilitates accurate voxel-level registration. The other stream learns the deep multi-scale residual representations to obtain robust recognition. We also factorize the 3D convolution to reduce the training parameters and enhance network efficiency.

Results: We validate the proposed method on a clinically acquired intra-patient abdominal CT-MRI dataset and a public inspiratory and expiratory thorax CT dataset. Experiments on both multimodal and unimodal registration demonstrate promising results compared to state-of-the-art approaches. 

Conclusion: By combining the high-resolution information and multi-scale representations in a highly interactive residual learning fashion, the proposed F3RNet can achieve accurate overall and local registration. The run time for registering a pair of images is less than 3 seconds using a GPU. In future works, we will investigate how to cost-effectively process high-resolution information and fuse multi-scale representations.

\end{abstract}

\section{Introduction}
In image-guided therapies (IGT), e.g., pre-operative planning, intervention and diagnosis, deformable image registration is the key to integrate complementary information contained in different time stamps or image modalities. Therefore, developing fast and accurate deformable image registration methods is beneficial for the performance of IGT.

Traditional registration methods such as Symmetric Normalization (SyN) \cite{Avants2008SymmetricDI} align a pair of images by iteratively minimizing the appearance dissimilarity under regularization constraints. Furthermore, Deeds \cite{heinrich2013mrf} utilizes discrete optimization, which shows promising results in abdominal registration \cite{xu2016evaluation}. However, solving a pairwise optimization is computationally intensive, resulting in slow speed in practice. Recently, due to the substantial improvement in computational efficiency over the traditional iterative registration, learning-based image registration approaches are becoming more prominent in task-specific and time-intensive applications \cite{fu2020review}. Most learning-based registration approaches use fully supervised \cite{fan2019birnet,lv2018supervised,chee2018airnet} or semi-supervised learning strategy \cite{hu2018weakly,ferrante2018weakly}, and heavily rely on ground-truth voxel correspondences and/or organ segmentation labels. Although these approaches struggle with imperfect ground-truth labels, they have made a significant impact on the field of deformable image registration. With the development of Spatial Transformer Network (STN) \cite{STN}, registration approaches that are based on unsupervised learning have also been introduced. For example, VoxelMorph \cite{VM2018} is a monumental unsupervised registration framework that focuses on registering brain images of the same modality (unimodal registration). By modifying VoxelMorph, researchers have further proposed more unsupervised unimodal registration approaches \cite{hu2019dual,de2019deep,kuang2019faim,ferrante2018unsupervised}. 


Most existing learning-based registration approaches use the so-called mono-stream ``high-to-low, low-to-high” network structure with augmented modules, e.g., skip-connection \cite{VM2018,ghosal2017deep}, multi-resolution fusion \cite{hu2019dual} and intermediate supervision \cite{li2018deepsuper}. This structure can significantly increase the size of the receptive field which is highly desirable for recognizing object information in images, but needs to recover the high-resolution information from the low-resolution representations. With increased receptive field sizes, these approaches prioritize overall registration accuracy, which is governed by the majority of easy-to-align regions, and overlook some severely deformed local regions. For example, livers with tumors usually have large local deformation due to progressed disease, and the deformations of the surrounding kidney and spleen are less significant. In a CT-to-MRI abdominal image registration, the aforementioned approaches are likely to estimate a deformation field that accurately registers kidney and spleen, yet perform poorly at local liver lobes alignment.

Besides, most of the image registration networks utilize 3D Convolutional Neural Networks (3D CNN) to exploit the semantic information in each CT/MRI slice and the spatial relationships across consecutive slices. It is understood that the training of 3D CNN is computationally expensive, and may lead to insufficient training due to the small number of clinical datasets.

To address the above problems, we propose a novel unsupervised \textbf{Full-Resolution Residual Registration Network (F3RNet)}, which is shown in \ref{fig1}(a). Distinct from the conventional mono-stream network structure, F3RNet consists of two parallel streams, namely ``Full-resolution stream'' and ``Multi-scale residual stream''. Inspired by the success of using a high-resolution stream in human pose estimation and image inpainting tasks \cite{sun2019deep,guo2019progressive,wang2020deep}, ``Full-resolution stream'' takes advantage of the detailed image information and facilitates accurate voxel-level registration. While the ``Multi-scale residual stream'' learns the deep multi-scale residual representations to robustly recognize corresponding organs in both images and guarantee a high overall registration accuracy. Using the Multi-scale Residual Block (MRB) modules, the network can progressively fuse information from the two parallel streams in a residual learning fashion \cite{he2016deep} to further boost the performance. In addition, we factorize the 3D convolution into two correlated 2D and 1D convolutions, thus effectively avoid over-parameterization \cite{szegedy2016rethinking}.

To the best of our knowledge, we are the first to incorporate full-resolution representations with multi-scale high-level representations in a residual learning fashion to boost deformable image registration performance. The main contributions of our work can be summarized as follows: 
\begin{itemize} 
\itemsep=0pt
\item Our approach can unite the strong capability of capturing deep multi-scale representations with precise full-resolution spatial localization of the anatomical structures by interactively combining two parallel streams via the proposed MRB module and the residual learning mechanism. By taking into account such full-resolution information, the registration network is more sensitive to the hard-to-align regions, and can provide better alignments for severely deformed local regions.


\item The factorization of 3D convolution can markedly reduce the training parameters and enhance the network efficiency.

\item We validate the proposed F3RNet on a clinically acquired intra-patient abdominal CT-MRI dataset and a public inspiratory and expiratory thoracic CT dataset. The experimental results on both multimodal and unimodal registration show that our method achieves superior performance over existing state-of-the-art traditional and learning-based methods.

\end{itemize}

The outline of the paper is as follows: Section \ref{methods} describes the details of our F3RNet, Section \ref{Experiments} presents the experimental details and registration results on both multimodal and unimodal datasets and Section \ref{Conclusions} will draw conclusions of the paper.


\section{Methods}
\label{methods}
\begin{figure}
	\centering
		\includegraphics[scale=0.4]{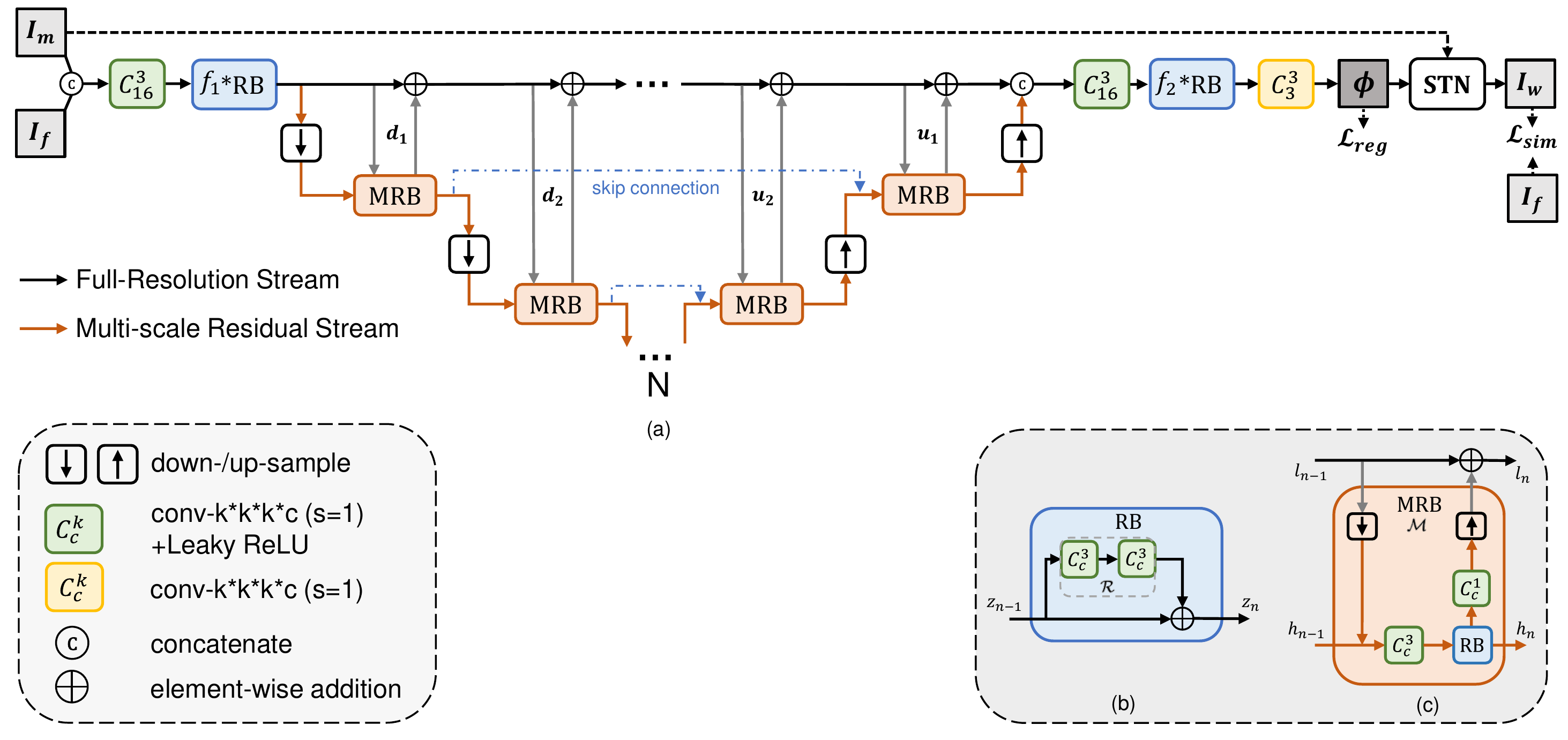}
	\caption{Illustration of Full-Resolution Residual Stream Network (F3RNet). (a) shows the overview of our F3RNet; (b) shows the residual block (RB); (c) shows the multi-resolution residual block (MRB). The network learns parameters for a dense deformation field $\phi$ that aligns the moving image $I_{m}$ to the fixed image $I_{f}$. $N$ denotes the minimum volume is $(1/2^{N})$ the size of the input images. }
	\label{fig1}
\end{figure}

Representing the moving image as $I_{m}$ and the fixed image as $ I_{f}$, medical image registration aims to estimate an optimal deformation field $\phi$ with three channels ($x$, $y$, $z$ displacements) that can align $I_{m}$ to $ I_{f}$. In this section, we present our Full-resolution Residual Registration Network (shown in Figure \ref{fig1}) firstly. Then, we describe the detailed structure of the designed Residual Block (RB) and Multi-scale Residual Block (MRB) respectively. The factorization of 3D convolution is presented in Section \ref{secF3D}, and the loss function of our network is described in Section \ref{loss}.
\subsection{Overview of the Network}
Distinct from the regular \textsl{high-to-low, low-to-high} one-pass network architecture, Full-Resolution Residual Registration Network (F3RNet) unifies two parallel streams:
\begin{itemize}
    \item \textbf{Full-resolution Stream}. Maintaining high-resolution features has demonstrated its superior performance for dense prediction \cite{sun2019deep,pohlen2017FRRN,wang2020deep,guo2019progressive}. The black line in Figure \ref{fig1}(a) indicates the data flow of the full-resolution stream. This stream first concatenates $ I_{m}$ and $ I_{f}$, followed by a 3D convolution and a series of Residual Blocks (RB, described in Section \ref{section:RB}). Then, the low-level features on this stream are successively computed by adding the residual from the other parallel stream. After that, the full-resolution stream reduces the number of channels via consecutive RBs and 3D convolutions step-by-step, and estimates the 3-channel deformation field $\phi$. Spatial Transformation Network (STN) \cite{STN} is applied to warp the moving image $ I_{m}$ with $\phi$, so that the similarity between the warped image $ I_{w}$ and fixed image $ I_{f}$ can be evaluated. This stream does not employ any downsampling operation, resulting in good boundary localization but poor performance in deep semantic recognition. As such, the hard-to-align regions are propagated throughout the stream. Specifically, the convolutions in the full-resolution stream are all with 16 channels in our experiments except for the final 3-channel convolution used to generate the deformation field.
    
    \item \textbf{Multi-scale Residual Stream}. The data flow of Multi-scale Residual Stream is depicted as the orange line in Figure \ref{fig1}(a). In contrast to the full-resolution stream, this stream is good at capturing high-level features that can improve the organ recognition performance. Specifically, successive pooling and convolution operations are leveraged to increase the receptive fields and enhance the robustness against small noises in the images. We also inherit the skip-connection design in regular \textsl{high-to-low, low-to-high} architecture that the feature spaces with same resolution are skip-connected by addition operation. Besides, with the help of our proposed Multi-scale Residual Blocks (MRB) that can simultaneously operate on both streams, the high-level features can directly interact with low-level features. The interior architecture of MRB is shown in Figure \ref{fig1}(c) with elaboration in Section \ref{sec.MRB}. In our experiments, we set $N$ to 4, which is the same as VoxelMorph \cite{VM2018}, denoting that the lowest resolution is $1/16$ of the original image. Specifically, at the resolution of $1/2$ and $1/4$ scale, the channel number of the feature map is set to 16. At the resolution of $1/8$ and $1/16$ scale, the number of feature channels becomes 32.
\end{itemize}

The information of the two distinct streams are automatically fused via residual learning \cite{he2016deep}. By repeatedly fusing the features between two streams via computing successive multi-scale residuals, the full-resolution representations become rich for the dense deformation field prediction. At the same time, richer low-level full-resolution information can in turn enhance the high-level multi-scale information.

\subsection{Residual Block (RB)}
\label{section:RB}
ResNets, proposed in \cite{he2016deep}, has demonstrated that residual learning can improve the training characteristics over traditional one-pass feed-forward learning. The interior architecture of the Residual Block (RB) is depicted in Figure \ref{fig1}(b). The output $z_{n}$ of the RB can be formulated as:
\begin{equation}
z_{n}=z_{n-1}+\mathcal{R}\left(z_{n-1}\right),
\end{equation}
where $\mathcal{R}$ represents the residual branch consisting of two 3D convolutions with a kernel size of $ 3 \times 3 \times 3$ followed by LeakyReLU activations. Instead of computing $z_{n}$ directly as in the traditional feed-forward network, the convolutional branch only needs to compute the residual $\mathcal{R}$ in this architecture. 

\subsection{Multi-scale Residual Block (MRB)}
\label{sec.MRB}
The Multi-scale Residual Block (MRB) follows the basic idea of Residual Block (RB) but elegantly achieves interaction between the full-resolution stream and multi-scale residual stream. An MRB consists of a series of pooling, 3D convolution and upsampling layers, as shown in Figure \ref{fig1}(c). Each MRB has two inputs, $l_{n-1}$ as full-resolution low-level features and $h_{n-1}$ as multi-resolution high-level features, and two corresponding outputs $l_{n}$ and $h_{n}$. Intuitively, denoting the entire MRB operation as $\mathcal{M}$, the output $l_{n}$ can be computed as:

\begin{equation}
l_{n}=l_{n-1}+\mathcal{M}\left(l_{n-1}, h_{n-1}\right).
\end{equation}

Specifically, first, the resolution of $l_{n-1}$ is reduced to that of $h_{n-1}$ by a pooling operation, followed by a feature map concatenation. Then, the concatenated feature map undergoes a 3D convolution with a kernel size of $3 \times 3 \times 3$, followed by a Residual Block (RB) with the same number of channels, and the output $h_{n}$ is connected to the next process of the multi-scale residual stream. Meanwhile, the output of the $3 \times 3 \times 3$ convolutional module adjusts the number of channels and the resolution to be consistent with $l_{n-1}$ through a $1 \times 1 \times 1$ convolutional bottleneck layer and an upsampling layer at the other end. By such a process, we can readily use addition operations to integrate the residuals learned in the MRB in the full-resolution stream, thus forming a dual-stream highly interactive residual module.	 

\subsection{Factorized 3D Convolution (F3D)}
\label{secF3D}

\begin{figure}
	\centering
		\includegraphics[scale=0.6]{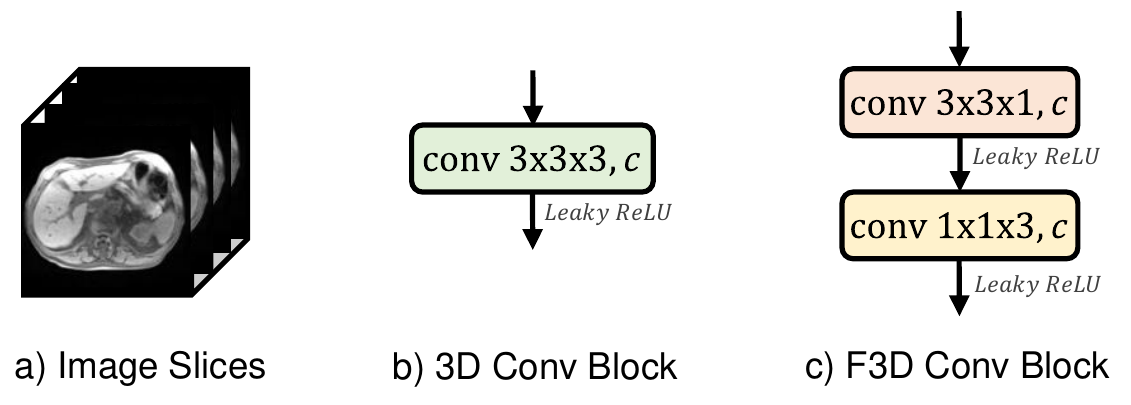}
	\caption{Illustration of a) 3D medical image scans, b) regular 3D convolution with kernel size of $3 \times 3 \times 3$, c) F3D convolution block.}
	\label{F3D}
\end{figure}

Most medical images, as shown in Figure \ref{F3D}(a), consist of 3D image stacks with the size of $W\times H\times D$, where $W$, $H$, $D$ represents the width, height, and the number of sequential slices. Inspired by the Inception \cite{szegedy2016rethinking} where large 2D convolution is factorized into two smaller ones, we factorize 3D convolution block for learning the volumetric representation. Specifically, suppose that we have a 3D convolution with kernel size of $3 \times 3 \times 3$ (Fig. \ref{F3D}(b)), it can be factorized into a $3 \times 3 \times 1$ convolution and a $1 \times 1 \times 3$ convolution in a cascaded fashion (Fig. \ref{F3D}(c)) to continuously capture dense 2D features in $W \times H$ slices with 1D attention weights that build sparse sequential relationships across adjacent slices. As such, the number of trainable parameters is reduced from $O(3^3=27)$ to $O(3 \times 3+3=12)$, where we can reduce the parameters by half.

However, it is noteworthy that the factorization is not totally equivalent to regular 3D convolution, and a further ablation study over factorized 3D convolution is presented in Section \ref{F3Dablation}.


\subsection{Loss Function}
\label{loss}
The loss function of our network consists of two components as shown in Eq.(\ref{mind_1}). The similarity loss $\mathcal{L}_{sim}$ penalizes the dissimilarity between the fixed image $I_{f}$ and the warped image $I_{w}(I_{m} \circ \phi)$. The deformation regularization $\mathcal{L}_{reg}$ adopts an L2-norm of the gradients of the final deformation field $\phi$ with a trade-off weight $\lambda$. We write the total loss as:

\begin{equation}
\label{mind_1}
\mathcal{L}(I_{m}, I_{f}, \phi)=\mathcal{L}_{sim}(I_{f}, I_{m} \circ \phi)+\lambda \mathcal{L}_{reg}(\phi).
\end{equation}

Specifically, \textsl{Modality Independent Neighborhood Descriptor (MIND)} \cite{mind} can be used to measure the similarity of both multimodal and unimodal images. MIND is a modality-invariant structural representation, and we can minimize the difference in the MIND features between the warped image $I_{w}(I_{m} \circ \phi)$ and the fixed image $I_{f}$ to effectively train the registration network. We define:


\begin{equation}
\mathcal{L}_{sim}\left(I_{f}, I_{m} \circ \phi \right)=\frac{1}{N|R|} \sum_{x}\left\|M I N D\left(I_{m} \circ \phi \right)-M I N D\left(I_{f}\right)\right\|_{1},
\end{equation}
where $N$ denotes the number of voxels in input images $I_{w}(I_{m} \circ \phi)$ and $I_{f}$, $R$ is a non-local region around voxel $x$.

\section{Experiments}
\label{Experiments}
\subsection{Dataset and Implementation}
In this work, we focus on the application of abdominal CT-MRI multimodal registration to improve the accuracy of Percutaneous Nephrolithotomy (PCNL). To further validate the effectiveness of our method, we also evaluate the proposed method on a public lung CT unimodal dataset \cite{lungdataset_2020}.

\begin{itemize}
\itemsep=0pt
\item \textsl{Abdominal CT-MRI dataset}: Under the IRB approved study, we obtained an proprietary intra-patient CT-MRI dataset containing paired CT and MR images from 50 patients. The liver, kidney and spleen in both CT and MRI were manually segmented for quantitative evaluation. Standard preprocessing steps, including affine spatial normalization, resampling and intensity normalization, were performed. The images were cropped into $144\times144\times128$ subvolume with 1mm isotropic voxels and divided into two groups for training (40 cases) and testing (10 cases). 

\item \textsl{Learn2reg 2020 Lung CT dataset} \cite{lungdataset_2020}: This dataset contains paired inspiratory and expiratory thorax CT images from 30 subjects (20 cases for training and 10 cases for testing). For all scans, the lung segmentation masks are provided for evaluation. Standard preprocessing steps, including affine spatial normalization and resampling, had been performed by the challenge organization. We further carried out intensity normalization and cropped images into $128\times128\times160$ subvolume.
\end{itemize}

The proposed method is implemented using Keras with the Tensorflow backend. We train the network on a NVIDIA Titan X (Pascal) GPU using Adam optimizer \cite{kingma2015adam} with a learning rate of 1e-5. The batch size is set to 1. As for the optimal trade-off weight $\lambda$, we conduct exhaustive grid search and select the value that achieves the highest average Dice scores of ROIs on hold-out test set.

\subsection{Measurement}
We evaluate the registration performance using a series of metrics for each method, mainly including Average Surface Distance (ASD) (lower is better) and the average Dice score (higher is better) between the segmentation masks of warped images and fixed images. Besides, the average number of voxels with non-positive Jacobian determinant ($|J_{\phi}| \leq 0$) in the deformation fields is counted for evaluating the diffeomorphism of the local deformation (lower is better). The standard deviation of the Jacobian determinant ($\sigma(|J_{\phi}|)$) is also calculated for evaluating the smoothness of transformations (lower is better).

\subsection{Experimental Results}
\subsubsection{Ablation Study of F3D Convolution}
\label{F3Dablation}
As mentioned in Section \ref{secF3D}, although convolution factorization can dramatically reduce the training parameters, it may not be totally equivalent to the regular 3D convolution in practice. Therefore, we investigate the different combinations of F3D convolution in our F3RNet. In our experiments, except for the final 3-channel 3D convolution used to generate the deformation field, other $3 \times 3 \times 3$ convolutions can be replaced. The variants of F3RNet are presented in Table \ref{tabf3d}. In particular, the number of parameters of F3RNet-w/ F3D is only 56.8\% of the original F3RNet. ``More MRBs" indicates that two extra MRBs are added at the lowest resolution path, which means that it is possible to use the reduced parameters to add more MRBs to enhance the network's learning capability.

\begin{table}[]
\centering
\caption{Different combinations of F3D convolution (\checkmark) in proposed F3RNet.}\label{tabf3d}      
\resizebox{8cm}{!}{
\begin{tabular}{c|c|c|c|c}
\hline
\multirow{2}{*}{Network} & \multirow{2}{*}{FR stream} & \multicolumn{2}{c|}{MS stream}                             & \multirow{2}{*}{More MRBs} \\ \cline{3-4}
                         &                            & \multicolumn{1}{l|}{Encoder} & \multicolumn{1}{l|}{Decoder} &                           \\ \hline
F3RNet-w/o F3D           &                            &                             &                             &                           \\
F3RNet-w/ F3D            & \checkmark                   & \checkmark                     & \checkmark                     &                           \\
F3RNet-Enc               &                            & \checkmark                     &                             &                           \\
F3RNet-Dec               &                            &                             & \checkmark                     &                           \\
F3RNet-FR                & \checkmark                    &                             &                             &                           \\
F3RNet-MS                &                            & \checkmark                     & \checkmark                     &                           \\
F3RNet-MRB               & \checkmark                    & \checkmark                     & \checkmark                     & \checkmark        \\\hline          
\end{tabular}}
\end{table}

\begin{figure}
	\centering
		\includegraphics[scale=0.8]{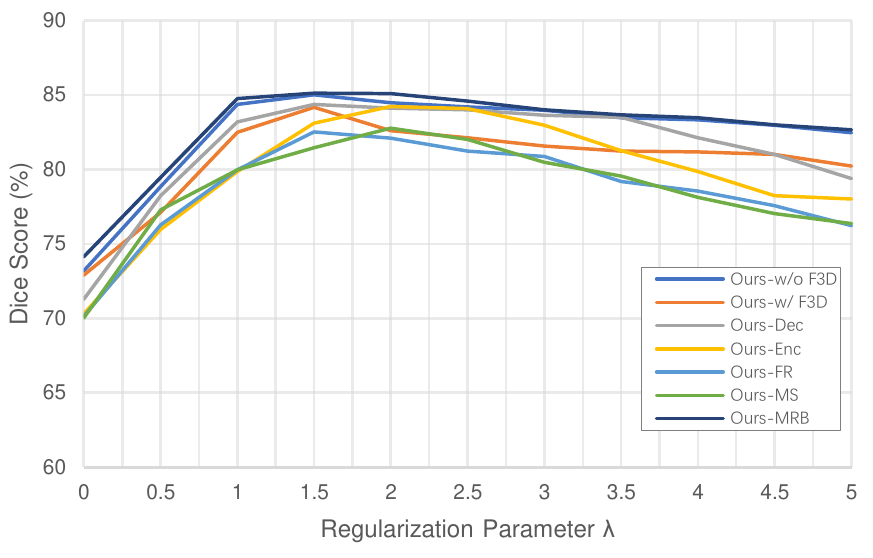}
	\caption{Results of varying the trade-off weight $\lambda$ on average Dice score of ROIs.}
	\label{reg_analysis}
\end{figure}

Figure \ref{reg_analysis} presents the average Dice scores of ROIs on the hold-out test set for varying values of the smoothing trade-off weight $\lambda$. The best Dice scores occur when $\lambda = 1.5$ for F3RNet-w/o F3D, F3RNet-w/ F3D, F3RNet-Dec, F3RNet-FR and F3RNet-MRB, and $\lambda = 2$ for F3RNet-Enc and F3RNet-MS. In particular, F3RNet-w/o F3D and F3RNet-MRB achieve better Dice scores than all other variants. Moreover, after achieving the best Dice scores at $\lambda = 1.5$, the results vary slowly over larger $\lambda$ for F3RNet-w/o F3D and F3RNet-MRB, showing that the two models are more robust to the choice of $\lambda$. 

\begin{figure}
	\centering
		\includegraphics[scale=0.4]{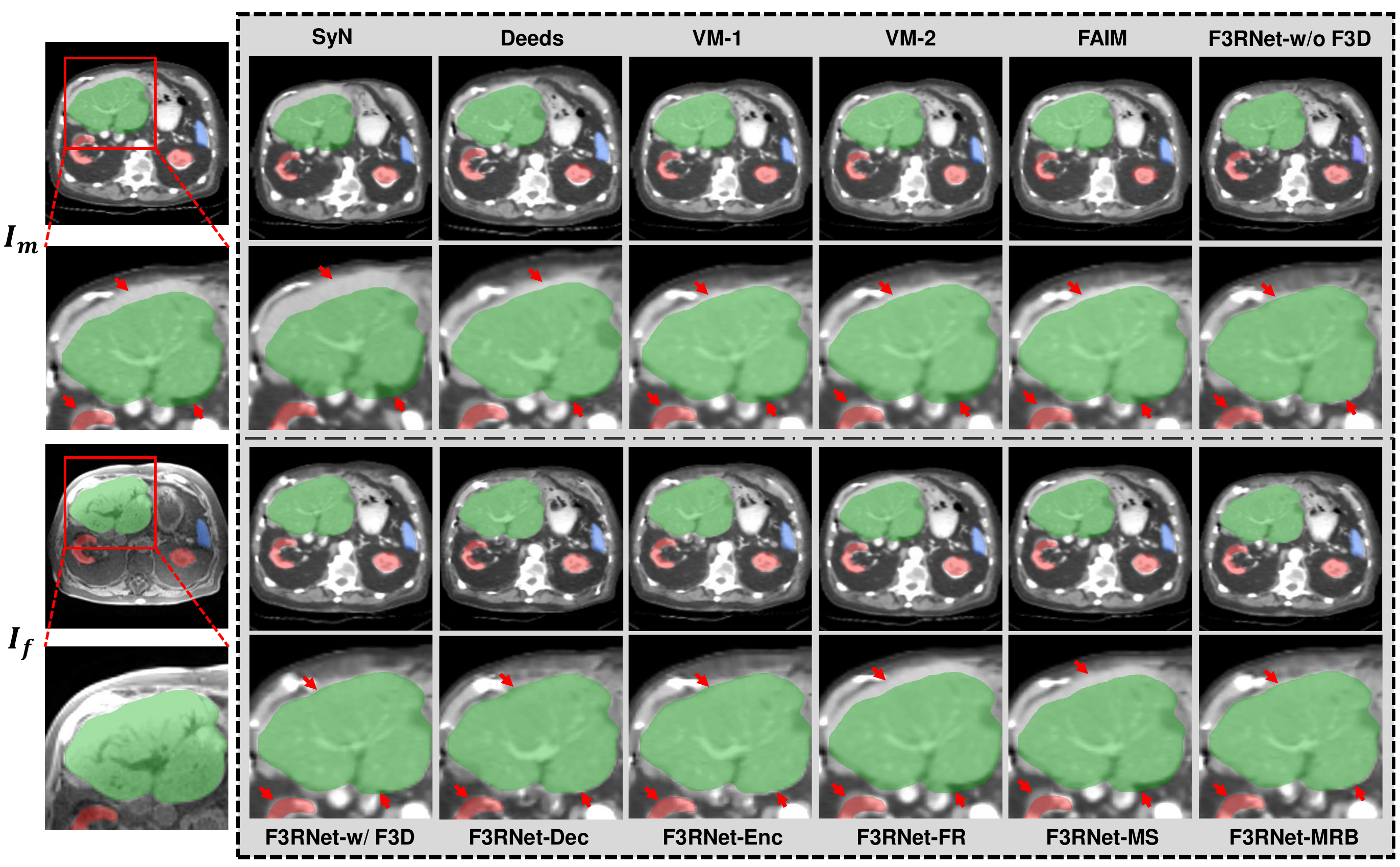}
	\caption{Visual results of an example for CT-to-MRI registration. Outside the grey box shows an example fixed MR image and a zoom-in region with the segmentation masks of the liver (green), kidney (red), and spleen (blue). The corresponding warped CT images and zoom-in regions for baselines and ablation study are presented in the grey box. A good registration will cause structures in warped images to close to the corresponding fixed segmentation masks. The red arrows indicate the registration of interest at the boundary of the organ.}
	\label{result1}
\end{figure}

Figure \ref{result1} shows visual results of warped images for the ablation analysis. We can firstly see that the original F3RNet (F3RNet-w/o F3D) can effectively register the multimodal images. If we replace all 3D convolutions with F3D (F3RNet-w/ F3D) or only replace the convolution in encoder and decoder (F3RNet-Enc and F3RNet-Dec), our methods can still effectively register the CT image but have slight performance degradation. Interestingly, if we replace the regular convolution on the entire multi-scale residual stream or full-resolution stream alone, this will cause the information of the two streams to not effectively interact and introduce noise, resulting in unstable performance and significant registration degradation. Therefore, if we use F3D to reduce the model parameters, the 3D convolution on both streams should be replaced at the same time. Further, we can use the reduced parameters to add more MRBs (F3RNet-MRB). From the visual results, it can be seen that the registration performance is either maintained or slightly improved.

\begin{table}
\caption{Average Dice scores and average ASD evaluations (mean$\pm$std) for CT-to-MRI registration of all baseline methods and F3RNet with different combinations of F3D. Best results are shown in bold. Besides, $|J_{\phi}| \leq 0$ (average number of folding voxels in the deformation fields) and $\sigma(|J_{\phi}|)$ (the smoothness of the deformation fields) are also provided.} \label{tab1}      
\resizebox{\textwidth}{!}{
\begin{tabular}{c|ccc|ccc|c|c|c}
\hline
\multirow{2}{*}{Methods} & \multicolumn{3}{c|}{ASD(mm)}                                                                                                           & \multicolumn{3}{c|}{Dice(\%)}   & \multirow{2}{*}{$\left |J_{\phi}\right| \leq 0 \; (\sigma(|J_{\phi}|))$}                                                             & \multirow{2}{*}{Params(M)} & \multirow{2}{*}{\begin{tabular}[c]{@{}c@{}}Times(s)\\ GPU/CPU\end{tabular}} \\ \cline{2-4} \cline{5-7}
                         & Liver                        & \multicolumn{1}{c}{Spleen}                      & \multicolumn{1}{c|}{Kidney}                      & \multicolumn{1}{c}{Liver}   & \multicolumn{1}{c}{Spleen}  & \multicolumn{1}{c|}{Kidney}  &                            &                                                                             \\\hline
Moving                   &4.95$\pm$0.82         & 1.97$\pm$0.52 &      2.01$\pm$0.36      & 77.18$\pm$4.13 & 78.24$\pm$3.21  & 80.14$\pm$3.17 &  -& -                          & -                         \\
SyN                      &       4.81$\pm$0.79       & 1.54$\pm$0.63 &      1.92$\pm$0.41         & 79.18$\pm$4.37 & 80.21$\pm$3.41  & 82.91$\pm$3.08 &  137.60(0.41)      & -                          & -/97                        \\
Deeds                      &      3.02$\pm$0.51       &   1.30$\pm$0.61   &       1.35$\pm$0.42       & 85.27$\pm$3.37  &  83.86$\pm$2.33  & 83.37$\pm$3.19 &   4.80(0.27)     &      -                     &      -/37                   \\
VM-1                     &   3.98$\pm$0.74           &1.42$\pm$0.54  &1.75$\pm$0.52               & 82.39$\pm$4.11 & 82.83$\pm$2.68 & 82.34$\pm$2.97 &   0.30(0.11)          & 0.260                      &1.32/21                      \\
VM-2                     & 3.92$\pm$0.53             &1.47$\pm$0.37  &1.72$\pm$0.49               & 84.17$\pm$3.57 & 82.76$\pm$2.98 & 83.51$\pm$3.36 &    1.10(0.15)      &0.300                      & 1.33/23                      \\
FAIM                     &3.88$\pm$0.73              &1.51$\pm$0.63  &1.66$\pm$0.41               & 84.51$\pm$3.76 & 81.99$\pm$2.84 & 83.12$\pm$3.24 &     0.00(0.12)      &0.217                      & 1.29/20                      \\\hline 
F3RNet-w/o F3D             & 2.19$\pm$0.37            &\textbf{1.28$\pm$0.39}  & 1.37$\pm$0.38               & 86.65$\pm$3.42 & 85.39$\pm$2.75 & \textbf{83.58$\pm$3.18} &      0.40(0.11)     &0.359                      & 2.31/26                      \\
F3RNet-w/ F3D              & 2.63$\pm$0.52              &1.36$\pm$0.86  &   1.41$\pm$0.72           & 85.16$\pm$4.19 & 84.39$\pm$2.23 & 82.98$\pm$3.43 &         0.80(0.09)    &0.204                      & 1.22/25                      \\
F3RNet-Dec                 & 2.77$\pm$0.69             &1.33$\pm$0.69  &   1.43$\pm$0.48            & 85.87$\pm$4.23 & 84.18$\pm$2.77 & 83.06$\pm$3.92 &        0.20(0.19)        & 0.335                      & 1.93/25                      \\
F3RNet-Enc                 &    2.73$\pm$0.47           & 1.39$\pm$0.52 &   1.38$\pm$0.36           & 85.32$\pm$4.35 & 84.27$\pm$3.35 & 83.11$\pm$3.28 &        0.50(0.17)       & 0.321                      & 1.79/23                      \\
F3RNet-FR                  &      3.82$\pm$0.59          & 1.53$\pm$0.39 &    1.56$\pm$0.43         & 81.43$\pm$4.27 & 82.97$\pm$3.18 & 83.13$\pm$3.52 &          0.70(0.24)     &0.348                      & 2.13/25                      \\
F3RNet-MS                  &  3.94$\pm$0.75            &1.51$\pm$0.44  &1.52$\pm$0.67               & 82.31$\pm$3.84 & 83.06$\pm$3.33 & 82.99$\pm$3.28 &     1.20(0.24)        & 0.243                      & 1.41/24                      \\
F3RNet-MRB                 & \textbf{2.17$\pm$0.46}              & 1.29$\pm$0.48 &   \textbf{1.34$\pm$0.26}           & \textbf{86.79$\pm$3.18} & \textbf{85.42$\pm$2.98} & 83.16$\pm$3.43 &    0.40(0.13)    &0.288                      & 1.42/25                     \\ \hline
\end{tabular}}
\end{table}

Table \ref{tab1} also provides the comprehensive quantitative results for all baseline methods and the variants of our F3RNet with different combinations of F3D. As for the results for ablation analysis, we can see that F3RNet-w/o F3D and F3RNet-MRB achieve the best performance. Specifically, with only 80.2\% parameters of F3RNet-w/o F3D, F3RNet-MRB achieves better ASD results in the liver and kidney registration than F3RNet-w/o F3D, while it also achieves better Dice score in liver and spleen registration with reasonable diffeomorphism and smoothness of the deformation fields. Meanwhile, consistent with the visual assessment, we can also see that F3RNet-FR and F3RNet-MS both yield significant performance degradation over ASD and Dice score as they cause the features of the two streams to be disjointed.

\subsubsection{Comparison with Baselines on abdominal CT-to-MRI registration}
To evaluate our proposed method, five open-source state-of-the-art baseline approaches are also compared, including two traditional methods SyN \cite{Avants2008SymmetricDI} with Mutual Information (MI) metric \cite{wells1996multi} and Deeds \cite{heinrich2013mrf} with five-levels of discrete optimization, and three unsupervised learning-based methods, marked as VoxelMorph-1 (VM-1) \cite{VM2018}, VoxelMorph-2 (VM-2) \cite{VM2018}, and FAIM \cite{kuang2019faim}. The three learning-based methods are initially proposed for unimodal registration, and we extend them for both multimodal and unimodal registration by using MIND-based similarity metric. We use the same test set to search for the best regularization weights and then set the weights to 1.5 for VM-1, VM-2 and FAIM. Other parameters, such as learning rate and batch size, remain the same as our method.

Figure \ref{result1} also illustrates the warped CT images produced by other baseline methods. As we have mentioned above, liver registration is much more challenging in the abdominal image registration task. From the results, we can see that the traditional method SyN fails to align the liver with large local deformation while Deeds performs much better. As for other deep learning methods, VM-1, VM-2, and FAIM achieve relatively satisfactory performance but still have considerable disagreements. Except for F3RNet-FR and F3RNet-MS, our methods have the most visually appealing boundary alignment, which demonstrates that our F3RNet can better register the hard-to-align regions. 

The quantitative results for the baseline methods are also presented in Table \ref{tab1}. Consistent with the visual results, the evaluations over ASD and Dice scores of our proposed methods are better than the traditional methods and other state-of-the-art unsupervised registration methods with reasonable quality of the deformation fields, except for F3RNet-FR and F3RNet-MS. Among the baseline methods, Deeds provides competitive results over SyN and other learning-based methods. Furthermore, the traditional methods take much more time (97s for SyN and 37s for Deeds) to register an image pair. In contrast, all deep learning methods can complete a registration task in 3 seconds with a GPU, making it appealing for image-guided therapies with intense time demand.

\subsubsection{Experiments on abdominal MRI-to-CT registration}
Among all the proposed networks for CT-to-MRI registration, F3RNet-w/o F3D and F3RNet-MRB provide superior results. To further validate the effectiveness of the two proposed methods, we also perform the MRI-to-CT registration in turn. The division of the dataset and the other training settings of the networks, e.g., regularization trade-off weights, etc, are consistent with the CT-to-MRI registration task. 

\begin{figure}
	\centering
		\includegraphics[scale=0.35]{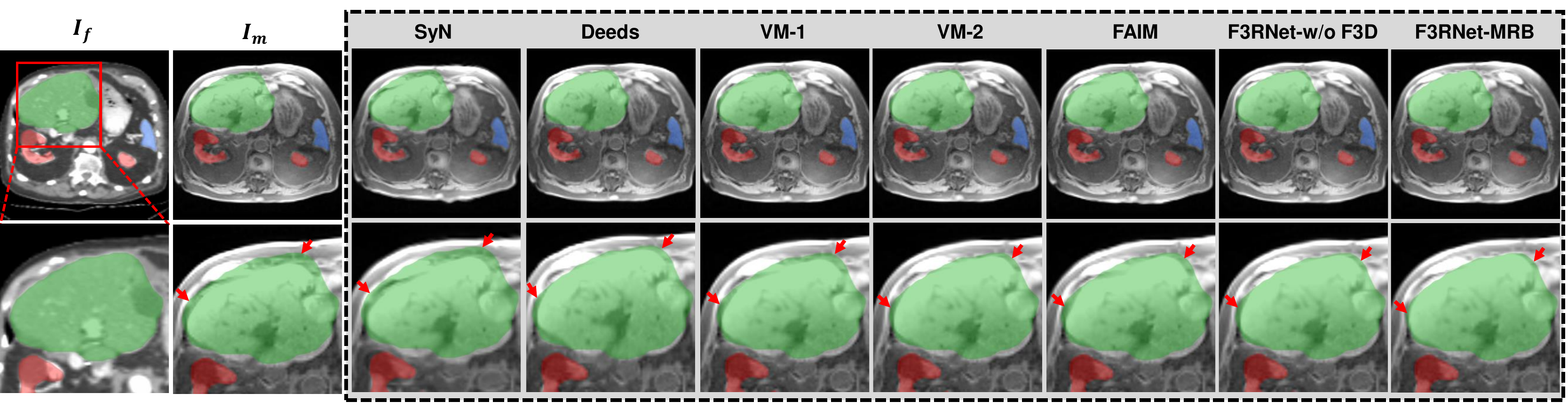}
	\caption{Visual results of an example for MRI-to-CT registration. Outside the grey box shows an example fixed CT image and a zoom-in region with the segmentation masks of the liver (green), kidney (red), and spleen (blue). The corresponding warped MR images and zoom-in regions for all methods are presented in the grey box. The red arrows indicate the registration of interest at the boundary of the organ.}
	\label{result2}
\end{figure}

The visualization of the registration results in Figure \ref{result2} shows that our methods, F3RNet-w/o F3D and F3RNet-MRB, achieve more accurate organ alignment than other traditional and deep learning approaches, especially for the liver. 

The quantitative evaluation of MRI-to-CT registration is summarized in Table \ref{tab2}. Our proposed methods achieve better results in terms of ASD and Dice scores than that of the traditional method and other state-of-the-art unsupervised learning registration methods. In particular, F3RNet-MRB achieves the best registration accuracy among all the methods with reasonably low $|J_{\phi}| \leq 0$ and $\sigma(|J_{\phi}|)$. 

\begin{table}[]
\caption{Average Dice scores and average ASD evaluations (mean$\pm$std) for MRI-to-CT registration. Best results are shown in bold. Besides, $|J_{\phi}| \leq 0$ (average number of folding voxels in the deformation fields) and $\sigma(|J_{\phi}|)$ (the smoothness of the deformation fields) are also provided.}\label{tab2}      
\resizebox{\textwidth}{!}{
\begin{tabular}{c|ccc|ccc|c}
\hline
\multirow{2}{*}{Methods} & \multicolumn{3}{c|}{ASD(mm)}                                                                                                 & \multicolumn{3}{c|}{Dice(\%)} &\multirow{2}{*}{$\left|J_{\phi}\right| \leq 0 \; (\sigma(|J_{\phi}|))$}                                                                                                      \\ \cline{2-4} \cline{5-7} 
                         & \multicolumn{1}{c}{Liver}               & \multicolumn{1}{c}{Spleen}              & \multicolumn{1}{c|}{Kidney}             & \multicolumn{1}{c}{Liver}               & \multicolumn{1}{c}{Spleen}                 & \multicolumn{1}{c|}{Kidney}              \\ \hline 
Moving                   & 4.95$\pm$0.82                           & 1.97$\pm$0.52                           & 2.01$\pm$0.36                           & 77.18$\pm$4.13                        & 78.24$\pm$3.21                              & 80.14$\pm$3.17  & -                      \\
SyN                      & 4.73$\pm$0.68                           & 1.63$\pm$0.57                           & 1.99$\pm$0.33                           & 78.36$\pm$4.67                        & 78.53$\pm$3.39                              & 81.38$\pm$2.98     &      89.2(0.33)             \\
Deeds                      &           3.58$\pm$0.53                 &          1.56$\pm$0.44                  &             1.84$\pm$0.40               &          83.59$\pm$3.87               &                 82.63$\pm$3.41              &    83.06$\pm$3.49     &      1.20(0.14)           \\
VM-1                     & 4.02$\pm$0.73                           & 1.61$\pm$0.64                           & 1.95$\pm$0.28                           & 81.34$\pm$4.06           & 80.72$\pm$3.02                           & 82.46$\pm$3.07        & 0.10(0.05)                \\
VM-2                     & 3.59$\pm$0.67                           & 1.53$\pm$0.52                           & 1.87$\pm$0.36                           & 83.28$\pm$4.03                           & 82.81$\pm$3.14                           & 83.37$\pm$2.83       &   0.00(0.03)                 \\
FAIM                     & 3.71$\pm$0.87                           & 1.51$\pm$0.63                           & 1.89$\pm$0.31                           & 84.33$\pm$3.64                           & 81.06$\pm$3.48                              & 83.44$\pm$2.92        &   0.00(0.04)                \\\hline 
F3RNet-w/o F3D             & 3.12$\pm$0.59                        & 1.43$\pm$0.59                           & 1.68$\pm$0.42                           & 85.75$\pm$4.11                        & 83.02$\pm$3.26                           & 84.07$\pm$3.04      &   0.00(0.06)                  \\
F3RNet-MRB                 & \textbf{3.04$\pm$0.65}              & \textbf{1.38$\pm$0.51} &   \textbf{1.67$\pm$0.35}           & \textbf{85.93$\pm$3.52} & \textbf{83.47$\pm$3.51} & \textbf{84.39$\pm$2.77}          &   0.00(0.04)            \\ \hline
\end{tabular}}
\end{table}

\subsubsection{Experiments on expiration-to-inspiration lung CT registration}

\begin{figure}
	\centering
		\includegraphics[scale=0.26]{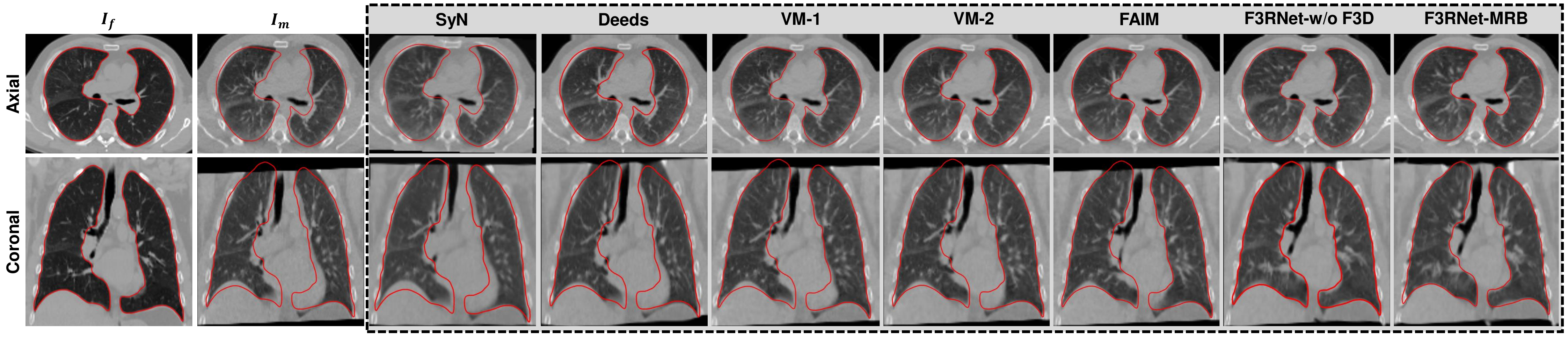}
	\caption{Visual results of an example for expiration-to-inspiration lung CT registration from both axial and coronal views. The red contours represent the lung segmentation of the fixed inspiratory CT image.}
	\label{lung_result}
\end{figure}

Apart from the large local deformation between expiratory and inspiratory lung CT images, another challenge of the Learn2Reg 2020 Lung CT dataset \cite{lungdataset_2020} is that the lungs are not fully visible in several expiratory scans as shown in $I_{m}$ of Figure \ref{lung_result}. In our experiment, MIND-based similarity metric \cite{mind} is still used to guide the network training. Empirically, the regularization weights are all set to 1.5 for VM-1, VM-2, FAIM and F3RNet. Other parameters, such as learning rate and batch size, remain the same as the aforementioned experiments.

\begin{table}[]
\centering
\caption{Average Dice scores and average ASD evaluations (mean$\pm$std) for lung CT registration. Best results are shown in bold. Besides, $|J_{\phi}| \leq 0$ (average number of folding voxels in the deformation fields) and $\sigma(|J_{\phi}|)$ (the smoothness of the deformation fields) are also provided.}\label{tab3}      
\resizebox{10cm}{!}{
\begin{tabular}{c|p{2.5cm}<{\centering}|p{2.5cm}<{\centering}|p{2.5cm}<{\centering}}
\hline
Methods        & ASD(mm) & Dice(\%) & $\left|J_{\phi}\right| \leq 0 \; (\sigma(|J_{\phi}|))$ \\\hline
Moving         &    2.51$\pm$0.83     &     86.37$\pm$3.45     &       -                                              \\
SyN(MI)        &    2.34$\pm$0.31     &    86.27$\pm$2.62      &      133.50(0.36)                                               \\
Deeds          &     2.35$\pm$0.72    &    87.28$\pm$2.78      &     7.50(0.11)                                                \\
VM-1           &   2.46$\pm$0.84      &    86.63$\pm$3.49      &      0.00(0.08)                                               \\
VM-2           &   2.38$\pm$0.82      &    87.05$\pm$3.39      &     1.20(0.09)                                                \\
FAIM           &   2.43$\pm$0.85      &   86.86$\pm$3.39       &    4.50(0.11)                                                 \\\hline
F3RNet-w/o F3D &     1.95$\pm$0.37    &    87.84$\pm$2.50      &   1.80(0.16)                                                  \\
F3RNet-MRB     &    \textbf{1.69$\pm$0.34}     &     \textbf{88.95$\pm$2.59}     &  2.10(0.13)    \\ \hline                                              
\end{tabular}}
\end{table}

We visualize an example of the registration results from both axial and coronal views in Figure \ref{lung_result}. Apparently, the proposed methods, F3RNet-w/o F3D and F3RNet-MRB, achieve more accurate lung alignment than other traditional and deep learning approaches, especially from the coronal view.

The quantitative evaluation of expiration-to-inspiration lung CT registration is summarized in Table \ref{tab3}. Our proposed methods achieve better results in terms of ASD and Dice scores than that of the traditional methods and other state-of-the-art unsupervised learning registration networks with reasonable tradeoff in the diffeomorphism and smoothness of the deformation fields. In particular, F3RNet-MRB achieves the best performance among all the methods.

\section{Conclusions}
\label{Conclusions}
In this work, we propose a novel unsupervised registration network, namely Full-Resolution Residual Registration Network (F3RNet), which takes advantage of full-resolution information, multi-scale fusion, deep residual learning framework and 3D convolution factorization, to improve the deformable registration performance. The experimental results on both multimodal and unimodal tasks indicate that our network can better register the hard-to-align region, yielding superior accuracy of registration. 
In our experiments, we found the current input size to be a compromise between image resolution and GPU memory limitation. Recently, the Laplacian Pyramid Image Registration Network (LapIRN) \cite{mok2020large} that includes three pyramid branches to register the image pairs at different resolutions with a coarse-to-fine optimization scheme is proposed, which brings promising enlightenment on improving multi-scale fusion-based registration. Future works will continuously focus on the lighter and more elegant ways to leverage high-resolution information and multi-scale fusion to cope with the large local deformation under limited GPU memory.

\begin{acknowledgements}
This project was supported by the National Institutes of Health (Grant No. R01EB025964, R01DK119269, and P41EB015898), the National Key R\&D Program of China (No. 2020AAA0108303), NSFC 41876098 and the Overseas Cooperation Research Fund of Tsinghua Shenzhen International Graduate School (Grant No. HW2018008).
\end{acknowledgements}

%
\section*{Conflict of interest}
Jayender Jagadeesan owns equity in Navigation Sciences, Inc. He is a co-inventor of a navigation device to assist surgeons in tumor excision that is licensed to Navigation Sciences. Dr.Jagadeesan’s interests were reviewed and are managed by BWH and Partners HealthCare in accordance with their conflict of interest policies.

\section*{Ethical approval}
All procedures performed in studies involving human participants were in accordance with the ethical standards of the institutional and/or national research committee and with the 1964 Helsinki declaration and its later amendments or comparable ethical standards.

\section*{Informed consent}
Informed consent was obtained from all individual participants included in the study.

\bibliographystyle{spmpsci}      
\bibliography{refs}   


\end{document}